\documentclass[twocolumn,showpacs,preprintnumbers,amsmath,amssymb]{revtex4}

\usepackage{graphicx}
\usepackage{dcolumn}
\usepackage{bm}

\newcommand{\be}{\begin{eqnarray}}
\newcommand{\ee}{\end{eqnarray}}

\begin{document}

\title{Markov processes follow from the principle of Maximum Caliber}

\author{Hao Ge$^{1,2}$}\email{gehao@fudan.edu.cn}
\author{Steve Press\'e$^{3}$}
\author{Kingshuk Ghosh$^{4}$}
\author{Ken A. Dill$^{5}$}
\affiliation{
$^1$ School of Mathematical Sciences\\
and Centre for Computational Systems Biology\\
Fudan University, Shanghai 200433, PRC\\
$^2$Department of Chemistry and Chemical Biology,\\
 Harvard University, Cambridge, MA 02138, USA.\\
$^3$ Department of Pharmaceutical Chemistry, University of California, San Francisco, CA 94158 \\
$^4$ Department of Physics and Astronomy, University of Denver, CO 80208\\
$^5$ Laufer Center for Physical and Quantitative Biology, Stony Brook University, NY 11794.
}

\date{\today}

\begin{abstract}
Markov models are widely used to describe processes of stochastic
dynamics. Here, we show that Markov models are a natural consequence
of the dynamical principle of Maximum Caliber.  First, we show that
when there are different possible dynamical trajectories in a
time-homogeneous process, then the only type of process that
maximizes the path entropy, for any given singlet statistics, is a
sequence of identical, independently distributed (i.i.d.) random
variables, which is the simplest Markov process.  If the data is in
the form of sequentially pairwise statistics, then maximizing the
caliber dictates that the process is Markovian with a uniform
initial distribution. Furthermore, if an initial non-uniform
dynamical distribution is known, or multiple trajectories are
conditioned on an initial state, then the Markov process is still
the only one that maximizes the caliber.  Second, given a model,
MaxCal can be used to compute the parameters of that model.  We show
that this procedure is equivalent to the maximum-likelihood method
of inference in the theory of statistics.
\end{abstract}

\maketitle

\section{Introduction}

E.T. Jaynes' principle of maximum entropy (MaxEnt) has wide
applications in engineering and science \cite{jaynes_book,
steinbach, gulldaniell}, and serves as one description of the
foundation for equilibrium statistical mechanics
\cite{jaynes_a,jaynes_b}.  In recent years, this principle has been
generalized for treating time-dependent statistical phenomena, and
is sometimes called the principle of maximum caliber (MaxCal)
\cite{jaynesmacro, dill_07,three-state-cycle,toggle,Otten2010}.

    Markov processes \cite{klchung_book}, whose transition densities
are described by Chapman-Kolmogorov equations \cite{vankampen}, are
a common starting point in modeling stochastic dynamics.   Here we
justify from a maximum entropy standpoint ``why start with a Markov
process?''

    In the present paper, we show that the application of
MaxCal to time-homogeneous data precisely yields a Markov
process. A Markov process model comes with a set of parameters,
i.e., the transition probabilities.  These are precisely related to the
Lagrange multipliers generated by the principle of MaxCal.
In determining these parameters, the MaxCal approach coincides with the method of maximum likelihood in the statistics of parameter estimation.


The theory of Maximum Caliber, in other words the idea of maximizing
path entropy subject to constraints, was first tackled in the
context of Markov processes by Filyukov and Karpov in 1967
\cite{fily}, as far as the current authors are aware. In this work,
it was assumed that a trajectory could be broken down as a Markov
chain. Motivated by arguments from Khinchin, Filyukov and
Karpov argued that maximizing the breadth of the trajectory ensemble
based on data provided on the Markov transition probabilities was a
recipe for inferring a trajectory ensemble consistent with
observation. Similar reasoning has been used in recent work
\cite{dill_07,three-state-cycle,toggle,Otten2010,Stock2008,Eric2011,Cecile2011,King2011}.

Here we work in reverse and ask, given particular constraints is it possible
to infer that the Markov model is the one that maximizes the entropy?
The answer will turn out to be yes as
maximum entropy is well known to set correlations between independent probabilities(or more generally, independent subsets of a map)
to zero unless data provides evidence for correlation. This derives from axioms used in obtaining the entropy formula, namely
``subset independence" \cite{liveseyskilling, shore}; a property originally enforced by Shannon through his  ``composition property"
as a logical consistency requirement for any inference\cite{shannon}.
Hence, given independent data on transition probabilities, a trajectory becomes the product of transition probabilities.
We verify the Markov process as following from consistency conditions for path probabilities as noted by a. Kolmogorov\cite{Kolmogorov} and maximum entropy.

\section{The Maximum-Caliber approach}

Suppose we have a stochastic process with a discrete state space
$S=\{1,2,\cdots,N\}$.

Consider its trajectories of length T, and denote the probability
for the trajectory $\{i_0,i_1,\cdots,i_T\}$ as $p_{i_0i_1\cdots
i_T}$ where $i_{n}$ is the state visited at time $n$. The path entropy, $S(T)$, is
\be
S(T)=-\sum_{i_0,i_1,\cdots,i_T}p_{i_0i_1\cdots
i_T}\log p_{i_0i_1\cdots i_T}.
\label{defS}
\ee

The above, Eq.(~\ref{defS}), is the correct measure which, when
maximized subject to constraints, yields the least biased set
$\{p_{i_0i_1\cdots i_T}\}$ consistent with observation \cite{shore, liveseyskilling}.
Furthermore, if $\{p_{i_0i_1\cdots i_T}\}$ is a path
probability, then it must satisfy certain logical consistency
requirements first established by A. Kolmogorov \cite{Kolmogorov}.

Specifically, for any $T$ equal $0$ or any positive integer, the probability
distribution has to satisfy
\begin{equation}
\sum_{i_{T+1}}p_{i_0\cdots i_Ti_{T+1}}=p_{i_0\cdots
i_T},\label{consist}
\end{equation}
i.e. the distribution of the process during the time interval
$\{0,1,2,\cdots,T\}$ should be the marginal distribution of the
process during the time interval $\{0,1,2,\cdots,T+1\}$.

We consider the three different types of constraints.\\

{\bf (1) Constraint on the mean of the singlet distribution.}  First, we suppose we only know the mean number of time intervals during which the system dwells in each state $m$, $A_{m}$, in a trajectory of
length T, where \be A_{m}=\sum_{i_0,i_1,\cdots,i_T}p_{i_0i_1\cdots
i_T}\sum_{k=0}^{T}\delta_{i_k,m}, \ee where $\delta_{i,j}=1$ when
integers $i=j$, and $0$ otherwise, $\sum_{k=0}^{T}\delta_{i_k,m}$ is
the number of time intervals during which state $m$ is occupied in
the trajectory $\{i_0,i_1,\cdots,i_T\}$, and $\sum_m A_m=T+1$. We
call this data singlet statistics.

Finally, we maximize Eq. (\ref{defS}) subject to constraints on our
singlet statistic imposed as Lagrange multipliers. That is, we
maximize the Caliber, $S(T)-\sum_{m}\lambda_{m}A_{m}(T)$ with
respect to $p_{i_0i_1\cdots i_T}$. We conclude that \be
p_{i_0i_1\cdots i_T}\propto
e^{-\sum_{m}\lambda_m\sum_{k=0}^{T}\delta_{i_k,m}}=\prod_{k=0}^T
e^{-\lambda_{i_k}}. \label{indep} \ee

Thus including normalization the probability now becomes \be
p_{i_0i_1\cdots i_T}&=&\frac{\prod_{k=0}^T
e^{-\lambda_{i_k}}}{\sum_{i_0i_1\cdots i_T}\prod_{k=0}^T
e^{-\lambda_{i_k}}}\nonumber\\
&=&\frac{\prod_{k=0}^T
e^{-\lambda_{i_k}}}{(\sum_{i=1}^Ne^{-\lambda_i})^{T+1}}=\prod_{k=0}^T
p_{i_k}(T),\ee

By definition
\be
A_{m}=\sum_{i_0,i_1,\cdots,i_T}p_{i_0i_1\cdots
i_T}\sum_{k=0}^{T}\delta_{i_k,m}=(T+1)p_m(T).\ee

As a final note on this example, we consider the result of imposing
the consistency condition,
 Eq. \ref{consist},  to the distribution obtained above. From this we obtain
\be \sum_{i_{T+1}}p_{i_0\cdots
i_Ti_{T+1}}=\prod_{k=0}^Tp_{i_k}(T+1), \ee but since
$\sum_{i_{T+1}}p_{i_0\cdots i_Ti_{T+1}}=p_{i_0i_1\cdots i_T}=
\prod_{k=0}^Tp_{i_k}(T)$, it then follows that \be
\prod_{k=0}^Tp_{i_k}(T+1)=\prod_{k=0}^Tp_{i_k}(T)
\label{consist1},\ee which gives \be p_{i_k}(T)=p_{i_k}(T+1)
\label{idendistr} \ee for any $i_{k}$. The proof is trivial if we
were now to change $i_T$ to another state $i_T^{'}$. This would give us \be
\prod_{k=0}^{T-1}p_{i_k}(T+1)\cdot
p_{i_T^{'}}(T+1)=\prod_{k=0}^{T-1}p_{i_k}(T)\cdot p_{i_T^{'}}(T) \ee
which, when compared with Eq. (\ref{consist1}), yields
$$\frac{p_{i_T^{'}}(T+1)}{p_{i_T}(T+1)}=\frac{p_{i_T^{'}}(T)}{p_{i_T}(T)}.$$
Then since both the summation of all $p_i(T+1)$ and $p_i(T)$ are
equal to one, hence Eq. (\ref{idendistr}) follows.

Under such constraints, MaxCal thus yields an identical, independent
distributed (i.i.d.) process. Eq. (\ref{indep}) is the statement of
independence and Eq. (\ref{idendistr}) is the statement of the
identical distributions.

{\bf (2) Constraint on pairwise statistics.}  Now we consider instead a situation in which the constraint, instead, is on the pairwise statistics for each step $m\rightarrow n$ in the time
period $[0,T]$, i.e.

\be
A_{mn}=\sum_{i_0,\cdots,i_T}p_{i_0\cdots
i_T}\sum_{k=0}^{T-1}\delta_{i_k,m}\delta_{i_{k+1},n}.
\ee
 Here
$\sum_{k=0}^{T-1}\delta_{i_k,m}\delta_{i_{k+1},n}$ is just the
number of occurrence of the transition $m\rightarrow n$, and
$\sum_{m,n} A_{mn}=T$.

Then maximizing Eq. (\ref{defS}) subject to constraints on the
pairwise statistic imposed using Lagrange multipliers, we conclude
that \be p_{i_0\cdots i_T}\propto
e^{-\sum_{m,n}\lambda_{mn}\sum_{k=0}^{T-1}\delta_{i_k,m}\delta_{i_{k+1},n}}
=\prod_{k=0}^{T-1} p_{i_ki_{k+1}}, \label{marko} \ee where
$p_{i_ki_{k+1}}=e^{-\lambda_{i_ki_{k+1}}}$.

Next, applying the consistency condition, Eq. (\ref{consist}), we
find by a method that is similar to the previous case (i.e. only replacing
$i_T$ with $i_T^{'}$ and considering its difference from the
original form) that
\be
 p_{i_ki_{k+1}}(T+1)= p_{i_ki_{k+1}}(T).
\ee That is, under constraints on transition, the only stochastic
process consistent with maximum caliber is the Markovian process
with uniform initial distribution. The Markovian property is
directly seen from the last proportionality in Eq. (\ref{marko}). We
use the equation above to justify that transition probabilities are
independent in the way that we used a similar expression to argue
independence of the i.i.d. process when we had singlet statistics.

{\bf (3) What if the initial distribution is not uniformly distributed?}  If the initial distribution is not uniformly
distributed, then the MaxCal method can still be used, but
with an entropy conditioned on an initial state as follows \be
S(T|i_0)=-\sum_{i_1,\cdots,i_T}p_{i_1\cdots i_T|i_0}\log
p_{i_1\cdots i_T|i_0}, \ee with constraints
\be
\langle\textrm{number of transitions}\ m \rightarrow n | i_0\rangle=
\nonumber\\
\sum_{i_1,\cdots,i_T}p_{i_1\cdots
i_T|i_0}\sum_{k=0}^{T-1}\delta_{i_k,m}\delta_{i_{k+1},n}=A_{mn}.
\ee

Then applying the Lagrange multiplier method, we would also conclude
that
$$p_{i_1\cdots i_T|i_0}=\prod_{k=0}^{T-1} p_{i_ki_{k+1}}.$$

Finally, we can easily derive a relation between $A_{mn}$, a property of the data, and
$\lambda_{mn}$, the Lagrange multiplier.
To do so, we define the {\it dynamical partition function}
\be
Q_d(T)=\sum_{i_1,\cdots,i_T} e^{-\sum_{m,n}\lambda_{mn}\sum_{k=0}^{T-1}\delta_{i_k,m}\delta_{i_{k+1},n}}.
\ee
Taking the derivative of this dynamical partition function gives the average number of time intervals of dwell, $A_{lp}=-\partial \log Q_d(T) / \partial \log \lambda_{lp}$ (resembling the way that taking derivatives of equilibrium partition functions yield equilibrium averages and higher cumulants).
When $T$ is large enough, we have $\sum_{l} A_{lp}(T)/T \rightarrow \pi_p$
where $\pi_p$ is the stationary distribution of state $p$.

\section{Data analysis procedure}

We have proved that the MaxCal approach gives a
Markov chain.  But, in practice, the constraints $A_{ij}(T)$ should be replaced by the
sample estimators $\hat{A}_{ij}(T)$.

The unconditioned trajectory probabilities are simply a reweighted
sum over the conditioned ones. By selecting the trajectories with
the same initial states from an unconditioned ensemble of
trajectories, one reproduces the distribution of the conditioned
trajectory ensemble. Conversely, by recombining the trajectory with
different initial states, one can obtain the distribution for the
unconditioned trajectory ensemble.

To be precise, if we had no information regarding the distribution
of initial conditions but if we had precise information regarding a
particular initial state for a specific measurement, then we could
only apply the conditional likelihood function
$$L(\{p_{ij}\})=\prod_{k=0}^{T-1} p_{i_ki_{k+1}}=\prod_{i,j} p_{ij}^{\hat{A}_{ij}(T)},$$
with constraints $\sum_{j}p_{ij}=1$ for each $j$.

Then applying the similar Langrange multiplier method like
maximizing the Caliber, we set
$H=L-\sum_i\lambda_i(\sum_{j}p_{ij}-1)$, where $H$ is the likelihood
subject to a constraint on normalization, and
$$\frac{\partial H}{\partial p_{ij}}=\hat{A}_{ij}(T)\frac{L(\{p_{ij}\})}{p_{ij}}-\lambda_i=0,$$
followed
$$p_{ij}\propto \frac{\hat{A}_{ij}(T)}{\lambda_i},$$
i.e.
$$p_{ij}=\frac{\hat{A}_{ij}(T)}{\sum_j \hat{A}_{ij}(T)}.$$

Knowing an initial distribution $p_i(0)$, simply forces us to add
some multiplicative constants in the likelihood function. The same
holds for multiple trajectories, except that we must group together
the trajectories with different initial states.

Thus, maximizing the conditional caliber in dynamical modeling is equivalent to maximizing the likelihood.

\section{Conclusions and Discussion}

Maximum caliber can be used both as a first principle from which to
derive stochastic dynamical models and also as a data analysis
method. We showed here that Markovian dynamics follows from MaxCal;
the Markov property is a natural consequence of maximizing a
dynamical entropy over trajectories for time-homogeneous processes.
Our treatment can be generalized to handle dynamical systems having
finite memory (time delays), corresponding to similar situations in
inference for time-series analysis \cite{brock}.


\section{Acknowledgement}
The authors would like to thank Prof. Hong Qian at the University of
Washington for very helpful discussions. We acknowledge NIH grant GM
34993 and 1R01GM090205. In addition, H. Ge acknowledges support by
the NSFC 10901040 and the specialized Research Fund for the Doctoral
Program of Higher Education (New Teachers) 20090071120003. This
article was completed when H. Ge visited the Xie group in the
Department of Chemistry and Chemical Biology, Harvard University. H.
Ge thanks Prof. Xiaoliang Sunney Xie and his department for their
support and hospitality.

\end{document}